\documentclass[12 pt, amsfonts,amsmath, amssymb,color]{article}
\usepackage{authblk}
\def\baselinestretch{1.0}
\usepackage{amsmath,amssymb,amsfonts,latexsym,float,graphics,epsfig}
\usepackage{subfig}
\usepackage{verbatim}
\usepackage{relsize}
\usepackage{hyperref}
\usepackage{graphicx}
\usepackage{bm}
\usepackage{epstopdf}
\usepackage{slashed}
\usepackage{color}
\usepackage{tikz-cd}
\usepackage[utf8]{inputenc}
\usepackage[T1]{fontenc}
\usepackage{fmtcount}

\def\be{\begin{equation}}
\def\ee{\end{equation}}
\def\bea{\begin{eqnarray}}
\def\eea{\end{eqnarray}}

\begin{document}

\renewcommand\theequation{\arabic{section}.\arabic{equation}}
\catcode`@=11 \@addtoreset{equation}{section}

\newtheorem{axiom}{Definition}[section]
\newtheorem{theorem}{Theorem}[section]
\newtheorem{axiom2}{Example}[section]
\newtheorem{lem}{Lemma}[section]
\newtheorem{prop}{Proposition}[section]
\newtheorem{cor}{Corollary}[section]

\newcommand{\ben}{\begin{equation*}}
\newcommand{\een}{\end{equation*}}

\renewcommand{\thefootnote}{\alph{footnote}}
\setcounter{footnote}{0}

\begin{titlepage}
\begin{center}
\renewcommand{\baselinestretch}{1.5}  

\vspace*{-0.5cm}

{\Large \bf{Hamiltonian thermodynamics on}}\\
 {\Large \bf{  symplectic manifolds}}

\vspace{9mm}
\renewcommand{\baselinestretch}{1} 

\centerline{\large{\bf Aritra Ghosh}$^\dagger$\footnote{{\fontsize{12pt}{14pt}\selectfont \textbf{Present Address:} School of Physics and Astronomy, Rochester Institute of Technology, Rochester, New York 14623, USA}} \large{\bf \&} \large{\bf E. Harikumar}$^\ddagger$}

\vspace{5mm}
\normalsize
\text{$^\dagger$School of Basic Sciences,} \\ 
\text{Indian Institute of Technology Bhubaneswar,}\\
\text{Jatni, Khurda, Odisha 752050, India}\\
\vspace{2mm}

\text{\textbf{Email:} aritraghosh500@gmail.com}

\vspace{7mm}

\text{$^\ddagger$School of Physics,} \\ 
\text{University of Hyderabad,}\\
\text{Central University P.O., Gachibowli,}\\
\text{Hyderabad, Telangana 500046, India}\\
\vspace{2mm}
\text{\textbf{Email:} eharikumar@uohyd.ac.in}

\vspace{1cm}

{\it With deep respect and admiration, we dedicate this\\ work to the memory of Professor A. P. Balachandran}

\vspace{0.4cm}

\begin{abstract}
We describe a symplectic approach towards thermodynamics in which thermodynamic transformations are described by (symplectic) Hamiltonian dynamics. Upon identifying the spaces of equilibrium states with Lagrangian submanifolds of a symplectic manifold, we present a Hamiltonian description of thermodynamic processes where the space of equilibrium states of a system in a certain ensemble is contained in the level set on which the Hamiltonian assumes a constant value. In particular, we work out two explicit examples involving the ideal gas and then describe a Hamiltonian approach towards constructing maps between related thermodynamic systems, e.g., the ideal (non-interacting) gas and interacting gases. Finally, we extend the theory of symplectic Hamiltonian dynamics to describe (a) the free expansion of the ideal gas which involves irreversible generation of entropy, and (b) a symplectic port-Hamiltonian framework for the ideal gas which is exemplified through two problems, namely, the problem of isothermal expansion against a piston and that of heat transfer between a heat bath and the gas via a thermal conductor. 
 \end{abstract}
\end{center}
\vspace*{0cm}

\end{titlepage}
\vspace*{0cm}

\makeatletter
\renewcommand{\thefootnote}{\arabic{footnote}}   
\renewcommand{\theHfootnote}{B-\arabic{footnote}} 
\makeatother
\setcounter{footnote}{0}

\section{Introduction}
Although the analogy between classical mechanics and thermodynamics has been known for a few decades \cite{herm,mrug,ajp,old0,old1,old2,old20,old22,old3}, the recent years have witnessed a renewed interest in the geometric approach to thermodynamics that uses geometric structures that arise naturally in classical mechanics (see for example, \cite{new1,new2,new3,new4,new5,new5_5,new5_6,new50,new6,new7,new8,new80,new9,new10,new11}). Focusing on equilibrium thermodynamics, the thermodynamic phase space, i.e., the space of all thermodynamic variables, is an odd-dimensional manifold on which one can define the so-called Gibbs one-form, naturally endowing it with the structure of a contact manifold \cite{herm,old1,old2,old20,old22,new1,new3,new4,new50,new7,new80,new11}. With this identification, it is observed that the equilibrium states of a thermodynamic system lie on certain special submanifolds of the ambient phase space -- such submanifolds are called Legendre submanifolds \cite{old2,new4,new6,new7,new9}. In this setting, different thermodynamic transformations can be described using Hamiltonian dynamics on contact manifolds by suitably choosing `contact' Hamiltonian functions that vanish on the space of equilibrium states; the latter condition is needed in order to ensure that the dynamical flow stays restricted to the space of equilibrium states (see \cite{old2,old22,new6,new7} for the details), i.e., the points on the flow trajectories are equilibrium states. The contact-geometric framework naturally generalizes to describe certain out-of-equilibrium situations with irreversibility as encountered in the general equation for nonequilibrium reversible-irreversible coupling (see \cite{new1,levinson} and references therein). 

\vspace{2mm}

Consider a hydrostatic system for which the first law of thermodynamics\footnote{In a strict physical sense, the first law of thermodynamics is only the statement of conservation of energy. The form (\ref{gibbs1}) suggested for a hydrostatic system has been supplemented by the fact that the incremental heat exchanged is given by $TdS$, typically true for reversible changes around equilibrium. In our formalism, the role of the statement (\ref{gibbs1}) and other similar statements will be to furnish constitutive relations such as (\ref{derivativesofE}) that are essential for the geometric identification. For the sake of simplicity, we will refer to (\ref{gibbs1}) as the first law of thermodynamics without strictly referring to its physical origin, and moreover, the thermodynamic states satisfying this, i.e., conforming to the constitutive relations (\ref{derivativesofE}), will be referred to as equilibrium states. While genuine equilibrium states conform to these requirements, there can be nonequilibrium states in certain systems (see for example, \cite{new9}) which also satisfy similar constitutive relations but we will not refer to them in this work.} is
\begin{equation}\label{gibbs1}
dE = TdS - PdV + \mu dN,
\end{equation} where the symbols have their usual meanings from thermodynamics. Notice that (\ref{gibbs1}) implies that $E = E(S,V,N)$, from which one obtains the relations
\begin{equation}\label{derivativesofE}
T = \bigg( \frac{\partial E}{\partial S}\bigg)_{V,N}, \quad P = - \bigg( \frac{\partial E}{\partial V}\bigg)_{S,N}, \quad \mu = \bigg( \frac{\partial E}{\partial N}\bigg)_{S,V}.
\end{equation}
Thus, the space of equilibrium states can be thought of as the three-dimensional space spanned by the variables $(S,V,N)$ such that the function $E = E(S,V,N)$ (called the fundamental equation) allows one to determine the rest of the thermodynamic variables $(T,-P,\mu)$ via the constitutive relations (\ref{derivativesofE}). The equilibrium states may be interpreted as being points on the above-mentioned three-dimensional submanifold of the thermodynamic phase space; the latter carries a natural contact structure (see \cite{mrug,old1,old2,old22} for details). Contact geometry \cite{Godbillon} is the odd-dimensional counterpart of symplectic geometry, which underlies conservative Hamiltonian mechanics \cite{arnold}. On a symplectic manifold, one can formulate Hamiltonian dynamics in a natural way which conserves the Hamiltonian (often the total energy). Pertaining to the geometric description of equilibrium thermodynamics, a natural question is whether equilibrium thermodynamics can be formulated on symplectic phase spaces? If so, can symplectic Hamiltonian dynamics describe in a consistent manner, thermodynamic transformations between equilibrium states analogous to contact Hamiltonian dynamics in the contact-geometric framework of thermodynamics? The answers to both the questions posed above are in the affirmative. 

\subsection{Motivation and results}
The purpose of this paper is to describe a geometric setting for thermodynamics and thermodynamic transformations between equilibrium states by using the framework of symplectic geometry rather than the commonly-employed framework of contact geometry (see \cite{new10} for a closely-related development; also see \cite{new5}). As we will discuss in detail, the equilibrium states of a thermodynamic system may be understood as being the points on a suitable `Lagrangian' submanifold of a symplectic manifold with descriptions associated with different statistical ensembles being related by Legendre transforms. 

\vspace{2mm}

As we shall demonstrate, the symplectic approach is not only capable of describing the geometry of equilibrium states, it also allows us to describe thermodynamic transformations using the familiar machinery of Hamiltonian dynamics. Moreover, it can be appropriately generalized to account for simple (phenomenological) irreversibilities. In particular, we will show that

\begin{enumerate}
\item A reversible thermodynamic transformation can be formally expressed as a Hamiltonian flow, restricted to the space of equilibrium states where the Hamiltonian is chosen to assume a constant value. This is the symplectic analogue of `Result VIII' proven in \cite{new6} in the context of contact geometry.

\item A nonsingular Legendre transform (corresponding to a change of ensemble) maps the thermodynamic evolution equations consistently between two ensembles. This is the symplectic analogue of a result proven in Section 4.1 of \cite{new4} in the context of contact geometry.

\item One may construct suitable Hamiltonian vector fields which map one thermodynamic system to another. We will furnish two novel examples in Secs. (\ref{24body}) and (\ref{RKsec}) which have not been discussed so far in the literature on geometrical thermodynamics. 

\item An irreversible process such as the free expansion of the ideal gas can be described using Hamiltonian dynamics and allows one to compute the entropy change consistent with traditional thermodynamic treatments. 

\item The framework of port-Hamiltonian systems \cite{port1,port2,port3,port4} can be seamlessly generalized to symplectic phase spaces describing thermodynamic systems (see also, \cite{new50,new80} for related developments focusing on the contact-geometric framework). We shall exemplify this by considering two problems: (a) the isothermal expansion of the ideal gas against a piston, leading to the explicit identification of thermal (heat bath) and mechanical (piston) ports, and (b) the problem of heat transfer between a heat bath and the ideal gas via a thermal conductor. 

\end{enumerate}

Thus, as far as the thermodynamics of equilibrium states is concerned, the symplectic framework can be treated as an alternative to the contact-geometric approach. The key motivation of this work is to present geometrical thermodynamics within the familiar framework of symplectic Hamiltonian dynamics. Since symplectic geometry forms the standard language of classical mechanics, this formulation makes the geometric approach to thermodynamics accessible to a wide audience while allowing direct use of the well-known Hamiltonian toolkit. We shall analyze several new explicit examples illustrating thermodynamic processes and mappings between systems. Let us begin with a brief review of the key concepts from symplectic geometry. 

\section{A brief review of symplectic geometry}\label{symplecticsec}

\subsection{Symplectic manifolds}
A symplectic manifold \cite{arnold} is the pair $(\mathcal{M},\omega)$, where $\mathcal{M}$ is a $2n$-dimensional smooth manifold and $\omega$ is a two-form that is both closed and non-degenerate, i.e., $d\omega = 0$ and $\omega^n \equiv \omega^{\wedge n} \neq 0$. Thus, $\omega^n$ describes a volume-form on $\mathcal{M}$. Darboux's theorem asserts that near a point, one can find a local system of coordinates $(q^i,p_i)$ such that 
\begin{equation}\label{omegadarboux}
\omega = dq^i \wedge dp_i.
\end{equation} 
In mechanics, the phase space of Hamiltonian systems is a cotangent bundle $\pi: T^* Q \rightarrow Q$, where $Q$ is the configuration space (the base manifold). Thus, if $q^i$ are the coordinates in an open subset of $Q$ with $p_i$ being the induced fiber coordinates, i.e., $\pi: (q^i,p_i) \rightarrow q^i$, then one can construct a tautological one-form which reads the following in these coordinates: 
\begin{equation}\label{thetadarboux}
\theta = p_i dq^i,
\end{equation} such that $\omega = - d\theta$. Such symplectic manifolds are called exact -- in this case, the symplectic two-form $\omega$ is an exact form. In general, however, the symplectic two-form may not be exact although it is closed by definition; in case of closed manifolds which are essentially compact and without a boundary, $\omega$ cannot be exact or else it will contradict Stokes' theorem. However, Darboux's theorem asserts that one can always define `local' coordinates in which the symplectic two-form looks like (\ref{omegadarboux}) or in other words, all $2n$-dimensional symplectic manifolds are locally isomorphic to $T^* \mathbb{R}^n$. 

\subsection{Hamiltonian dynamics}
Let us now describe Hamiltonian dynamics. Consider a symplectic manifold $(\mathcal{M},\omega)$. The non-degeneracy of $\omega$ allows the definition of a vector-bundle isomorphism between the tangent and cotangent bundles of $\mathcal{M}$ as
\begin{equation}\label{dHsymplecticdef}
\iota_{X_H}\omega = dH,
\end{equation} where $H \in C^\infty(\mathcal{M},\mathbb{R})$ is called the Hamiltonian function. It can now be verified by explicit calculation that in local (Darboux) coordinates, the vector field $X_H$ takes the following appearance so as to satisfy the condition (\ref{dHsymplecticdef}) along with (\ref{omegadarboux}):
\begin{equation}
X_H = \frac{\partial H}{\partial p_i} \frac{\partial}{\partial q^i} - \frac{\partial H}{\partial q^i} \frac{\partial}{\partial p_i}.
\end{equation} 
For any function $f \in C^\infty(\mathcal{M}, \mathbb{R})$, one has $\dot{f} \equiv X_H(f)$ and which consistently gives $X_H(H) = 0$, indicating the conservation of the Hamiltonian function. We specifically get
\begin{equation}\label{eom}
\dot{q}^i \equiv X_H(q^i) = \frac{\partial H}{\partial p_i}, \quad \quad \dot{p}_i \equiv X_H(p_i) = -\frac{\partial H}{\partial q^i},
\end{equation} i.e., the integral curves of the vector field $X_H$ satisfy the Hamilton's equations. We shall therefore refer to $X_H$ as the Hamiltonian vector field. An interesting consequence of (\ref{dHsymplecticdef}) is that the Lie derivative of the symplectic two-form with respect to a Hamiltonian vector field vanishes, i.e.,
\begin{equation}
\pounds_{X_H} \omega = \iota_{X_H} d\omega  + d(\iota_{X_H} \omega) = 0,
\end{equation} where the first term vanishes because $\omega$ is closed while the second term vanishes upon using (\ref{dHsymplecticdef}) because $d^2 = 0$. This implies that $\omega$ (and hence the volume-form $\omega^n$) is conserved under the flow of $X_H$, a result known as Liouville's theorem. 

\subsection{Lagrangian submanifolds}
Let us now define Lagrangian submanifolds which will be of chief interest in the context of thermodynamics. Consider a submanifold $L \subset \mathcal{M}$ such that $\phi: L \hookrightarrow \mathcal{M}$ is the relevant inclusion map. Then, if $\phi^* \omega = 0$, the submanifold $L$ is said to be an isotropic submanifold of $(\mathcal{M},\omega)$. Resorting to local (Darboux) coordinates as in (\ref{omegadarboux}), one finds that an isotropic submanifold should not possess a pair of $q^i$ and $p_i$ for the same $i$. Then it is somewhat intuitive that the dimensionality of an isotropic submanifold is less than or equal to half the dimensionality of the symplectic manifold, i.e., ${\rm dim~}L \leq n$ if ${\rm dim~}\mathcal{M} = 2n$. If $L$ is a maximal-dimensional isotropic submanifold, i.e., it is $n$-dimensional and satisfies the condition $\omega|_{L} = 0$, then it is called a Lagrangian submanifold. It turns out that all the Lagrangian submanifolds are $n$-dimensional and that their local structure is determined by the condition $\theta|_L = d\mathcal{F}$, for some suitable function $\mathcal{F}$. Using the expression (\ref{thetadarboux}), we get
\begin{equation}\label{Lagrangianlocal}
d\mathcal{F}(q^i) = p_i dq^i \quad \quad \implies \quad \quad p_i = \frac{\partial \mathcal{F}(q^i)}{\partial q^i},
\end{equation} where $i \in \{1,2,\cdots,n\}$. The function $\mathcal{F}(q^i)$ is termed as the generator of the Lagrangian submanifold. Notice that one can perform a Legendre transform on $\mathcal{F}(q^i)$ to get a function which generates a different Lagrangian submanifold; the two Lagrangian submanifolds are diffeomorphic if the Legendre transform connecting the two generators is nonsingular \cite{arnold} (see \cite{new4} for some related discussion). 

\section{Symplectic geometry of thermodynamics}\label{geomsec}
In this section, we shall describe the rich geometric structure of thermodynamics within the framework of symplectic geometry, essentially following up on the closely-related developments reported earlier \cite{new5,new10}. The starting point is the first law of thermodynamics which involves the fundamental equation. If $q^i$ with $i \in \{1,2,\cdots,n\}$ denote the thermodynamic variables which are the arguments of the fundamental equation $\Phi = \Phi(q^i)$, then the first law of thermodynamics is summarized by\footnote{As mentioned in footnote-(1), we will refer to statements like (\ref{Firstlawgeneral}) as the first law of thermodynamics. In our construction, their role is to supply the constitutive relations allowing the restriction to Lagrangian submanifolds.} 
\begin{equation}\label{Firstlawgeneral}
d\Phi(q^i) = p_i dq^i, \quad \quad p_i = \frac{\partial \Phi(q^i)}{\partial q^i},
\end{equation} where $p_i$ are the `thermodynamic' conjugate variables. Thus, comparing with the relation (\ref{Lagrangianlocal}), one may conclude that (\ref{Firstlawgeneral}) describes a Lagrangian submanifold of a symplectic manifold with the symplectic two-form $\omega = dq^i \wedge dp_i$. The coordinates on this Lagrangian submanifold are $q^i$ and the variables $(q^i,p_i)$ may be understood as Darboux coordinates near a point. The fundamental equation $\Phi = \Phi(q^i)$ generates this Lagrangian submanifold which physically corresponds to the space of equilibrium states. Thus, let us make the following proposition: 

\begin{prop}
Thermodynamic equilibrium states are described (locally) by points on a Lagrangian submanifold of a symplectic manifold with the thermodynamic potential as the generator of this submanifold. 
\end{prop}

Alternatively, one can begin with a description where one has a certain finite number of externally-controllable thermodynamic variables $q^i$ which may be thought of as being the local coordinates in some open subset of $\mathbb{R}^n$. Then, given an appropriate potential function $\Phi = \Phi(q^i)$ (typically dictated by statistical mechanics), the space of equilibrium states can be interpreted as a Lagrangian submanifold of a symplectic manifold which locally appears as $T^* \mathbb{R}^n$ with the symplectic two-form $\omega = dq^i \wedge dp_i$ such that $p_i$ are the corresponding induced fiber coordinates on the cotangent bundle. We will therefore define a thermodynamic system as follows: 

\begin{axiom}\label{definition}
A thermodynamic system is the triple $(\mathcal{M}, \omega, \mathcal{E})$, where $(\mathcal{M},\omega)$ is a symplectic manifold and $\mathcal{E} \subset \mathcal{M}$ is a Lagrangian submanifold. The local structure of $\mathcal{E}$ is dictated by a thermodynamic potential $\Phi \in C^\infty(\mathcal{E}, \mathbb{R})$ which satisfies the first law of thermodynamics. 
\end{axiom}

\subsection{Change of representation}
Consider a thermodynamic system with the fundamental equation $\Phi = \Phi(q^i)$ which satisfies the first law of thermodynamics (\ref{Firstlawgeneral}). Consider some specific $l \in \{1,2,\cdots, n\}$. If $p_l \neq 0$, (\ref{Firstlawgeneral}) may be rewritten as
\begin{equation}\label{flnew}
dq^l + \bigg(\frac{p_{i'}}{p_l}\bigg) dq^{i'} - \frac{d\Phi}{p_l} = 0,
\end{equation} where $i' \in \{ 1,2,\cdots, l-1, l+1, \cdots , n\}$. Since $p_l = \frac{\partial \Phi}{\partial q^l} \neq 0$, the function $\Phi = \Phi(q^1,q^2, \cdots, q^l, \cdots q^n)$ can be solved in favor of $q^l$ using the implicit-function theorem to write 
\begin{equation}
q^l = q^l (q^1, q^2, \cdots, q^{l-1}, q^{l+1}, \cdots, q^n, \Phi).
\end{equation} Thus, referring to (\ref{flnew}), we have a new first law of thermodynamics in which $q^l = q^l (\cdots)$ plays the role of the thermodynamic potential. We shall refer to (\ref{Firstlawgeneral}) and (\ref{flnew}) as two different representations of the same thermodynamic system. The reader is referred to Appendix (\ref{appA}) for a concrete example. 

\subsection{Legendre transforms}
In thermodynamics, one often encounters a change of ensemble in which one performs a Legendre transform on the thermodynamic potential so that it becomes a function of a different set of variables. For example, the internal energy of a hydrostatic system is a function of the entropy, volume, and number of particles but a Legendre transform takes it to the Helmholtz free energy which is a function of the temperature, volume, and number of particles. Notice that the two descriptions correspond, respectively, to the microcanonical and canonical ensembles. Now, for a thermodynamic system in the sense as described in Definition \ref{definition}, a (partial) Legendre transform may be expressed in the local coordinates as $\Psi(q^j,p_k) = \Phi(q^i) - p_k q^k$, where $i \in \{1,2,\cdots, n\}$, $j \in J$, $k \in K$, with $J \cup K = \{1,2,\cdots,n\}$ and $J \cap K = \{\}$. The first law of thermodynamics gets modified to
\begin{equation}
d\Psi(q^j,p_k) = p_j dq^j - q^k dp_k, \quad \quad p_j = \frac{\partial \Psi(q^j,p_k)}{\partial q^j}, \quad \quad q^k = - \frac{\partial \Psi(q^j,p_k)}{\partial p_k}. 
\end{equation}
Thus, the change of ensemble corresponds to constructing a map between two Lagrangian submanifolds of the ambient symplectic manifold. It is therefore clear that this mapping is a bijection (at least locally) if and only if the Legendre transform is regular, i.e., if the Hessian matrix of $\Phi(q^i)$ with respect to the arguments $q^k$ where $ k \in K$ is non-singular everywhere\footnote{This follows from the fact that if the above-mentioned Hessian is non-singular, the relations $p_k (q^i) = \frac{\partial \Phi(q^i)}{\partial q^k}$ can be solved in favor of the $q^k$'s.}. If the Legendre transform is regular, it acts as a local diffeomorphism\footnote{A simple way to see this is that a nonsingular Legendre transform is its own inverse, i.e., it is an involution map.} between a pair of Lagrangian submanifolds. See \cite{new4} for a detailed discussion on Legendre transforms but in the context of contact geometry as applied to thermodynamics. The reader is referred to Appendix (\ref{appB}) for some concrete examples. 

\subsection{Hamiltonian description of thermodynamic processes}\label{hamsec}
Since one can describe Hamiltonian dynamics on symplectic manifolds, this will allow us to describe the evolution of thermodynamic variables on the symplectic phase space on which the local (Darboux) coordinates are the conjugate variables appearing in thermodynamics. Such a construction shall describe the evolution of thermodynamic variables which may constitute a thermodynamic process of the system under consideration, e.g., isochoric transformation of an ideal gas. However, when one describes thermodynamic processes associated with a particular system, one must ensure that the Hamiltonian flow should be restricted to the space of equilibrium states -- otherwise, even if one picks an initial point on the phase trajectory to be an equilibrium state of the system, the trajectory may subsequently pass through points in the phase space that are not equilibrium states of the system under consideration. 

\vspace{2mm}

In other words, in order to provide a Hamiltonian description of a thermodynamic process, the space of equilibrium states should be invariant to the flow of the Hamiltonian vector field. Since Hamiltonian dynamics on symplectic manifolds conserves the Hamiltonian function, the level sets where the Hamiltonian assumes a constant value are invariant to the flow of the corresponding Hamiltonian vector field. Thus, given a thermodynamic system $(\mathcal{M}, \omega, \mathcal{E})$, one must choose a Hamiltonian $H$ to describe a certain thermodynamic process in such a way that the Hamiltonian takes a constant value on the space of equilibrium states, i.e., the space of equilibrium states is contained within the $H =$ constant surface; moreover, $X_H$ must be tangent to $\mathcal{E}$. More formally, one may furnish the following definition of a thermodynamic processes of a given system: 

\begin{axiom}
A thermodynamic process is the quadruple $(\mathcal{M}, \omega, \mathcal{E}, H)$, where $(\mathcal{M}, \omega, \mathcal{E})$ is a thermodynamic system and $H \in C^\infty(\mathcal{M}, \mathbb{R})$ is a Hamiltonian function such that $\mathcal{E} \subset H^{-1} (\Lambda)$ with $\Lambda \in \mathbb{R}$ being a constant and $X_H$ is tangent to $\mathcal{E}$. 
\end{axiom}

\begin{theorem}\label{theorem1}
Consider a thermodynamic system $(\mathcal{M},\omega,\mathcal{E})$ with the potential function $\Phi \in C^\infty(\mathcal{E},\mathbb{R})$. In Darboux coordinates $(q^i,p_i)$, let $q^i$ be the independent thermodynamic variables on $\mathcal{E}$, i.e., we may write $\Phi = \Phi(q^i)$. Then, given a thermodynamic process $\dot{q}^i = X^i(q^i)$ on the space of equilibrium states with $X^i(q^i)$ being suitable functions on $\mathcal{E}$, it can be described by the following choice of Hamiltonian: 
\begin{equation}\label{Hchoicegeneral}
H(q^i,p_i) = \bigg(p_i - \frac{\partial \Phi(q^i)}{\partial q^i} \bigg) X^i(q^i) + \Lambda,
\end{equation} where $\Lambda \in \mathbb{R}$ is a constant.
\end{theorem}

\textit{Proof --} Choosing a Hamiltonian that looks like (\ref{Hchoicegeneral}), Hamilton's equations for the variables $q^i$ are obtained to be 
\begin{equation}
\dot{q}^i = \frac{\partial H}{\partial p_i} = X^i(q^i). 
\end{equation}
Noting that on the space of equilibrium states $\mathcal{E}$ of the system, one has $p_i = \frac{\partial \Phi(q^i)}{\partial q^i}$, one finds that the restriction of $H$ to $\mathcal{E}$ is a constant, i.e., $H|_\mathcal{E} = \Lambda$. This fact, however, does not immediately guarantee the tangency of  $X_H $ to  $\mathcal{E} $ as the Hamiltonian flow with an initial point being on  $\mathcal{E} $ may flow to points that lie within  $H^{-1}(\Lambda) $ but outside $\mathcal{E} $. In order to check whether the space of equilibrium states is an invariant set in itself, the following condition must be satisfied for  $t > 0 $:
\begin{equation}
X_H\bigg(p_i - \frac{\partial \Phi(q^i)}{\partial q^i}\bigg)\bigg|_\mathcal{E} = 0,
\end{equation}
provided  $p_i - \frac{\partial \Phi(q^i)}{\partial q^i} = 0 $ at  $t=0 $. A direct calculation reveals that
\begin{equation}
X_H(p_i)-X_H\bigg(\frac{\partial \Phi(q^i)}{\partial q^i}\bigg) =-\bigg(p_j-\frac{\partial \Phi(q^i)}{\partial q^j}\bigg)\frac{\partial X^j(q^i)}{\partial q^i}.
\end{equation}
On  $\mathcal{E} $, the right-hand side vanishes due to the constitutive relations. So if the constitutive relations hold at  $t=0 $, they are preserved for all $t$. Thus, on the space of equilibrium states, the Hamiltonian (\ref{Hchoicegeneral}) describes the desired thermodynamic process with the corresponding flow preserving the constitutive relations.

\vspace{2mm}

Let us recall that one may perform Legendre transforms which can map different Lagrangian submanifolds to one another, each corresponding to a different ensemble. It follows that if the Legendre transform is regular, i.e., nonsingular, then the thermodynamic process is preserved. This can be summarized as follows: 

\begin{theorem}\label{theorem2}
Let $(\mathcal{M}, \omega, \mathcal{E}, H)$ be a thermodynamic process of a system on its space of equilibrium states $\mathcal{E}$ in a certain ensemble. Suppose we perform a Legendre transform to convert to a different ensemble, i.e., we have the map $\psi: \mathcal{E} \rightarrow \overline{\mathcal{E}}$, where $\overline{\mathcal{E}}$ is a different Lagrangian submanifold. Then, if the Legendre transform is regular, it preserves the thermodynamic evolution described by $H \in C^\infty(\mathcal{M},\mathbb{R})$. 
\end{theorem}

\textit{Proof --} Consider a thermodynamic system $(\mathcal{M}, \omega, \mathcal{E}, H)$ with the potential function $\Phi \in C^\infty(\mathcal{E},\mathbb{R})$. In Darboux coordinates $(q^i,p_i)$, let $q^i$ be the independent thermodynamic variables on $\mathcal{E}$, i.e., we may write $\Phi = \Phi(q^i)$. Now consider an arbitrary Legendre transform to define a new potential function $\Psi(q^j,p_k) = \Phi(q^i) - p_k q^k$, where $i \in \{1,2,\cdots, n\}$, $j \in J$, $k \in K$, with $J \cup K = \{1,2,\cdots,n\}$ and $J \cap K = \{\}$. The new potential describes a Lagrangian submanifold $\overline{\mathcal{E}}$ such that the Legendre transform forms a map between the Lagrangian submanifolds $\psi:\mathcal{E} \rightarrow \overline{\mathcal{E}}$. 

\vspace{2mm} 

Given a thermodynamic process $\dot{q}^i = X^i$ on $\mathcal{E}$, one can construct an infinitesimal vector field $X$ which is tangent to the trajectories such that one can write $X = X^i \frac{\partial}{\partial q^i}$. Then, the pushforward of this vector field under the map $\psi$, i.e., $\psi_*:T\mathcal{E} \rightarrow T\overline{\mathcal{E}}$ can be computed in Darboux coordinates to be 
\begin{eqnarray}
\psi_* X &=& X^j \frac{\partial}{\partial q^j} + X^{k'} \frac{\partial p_{k}}{\partial q^{k'}} \frac{\partial}{\partial p_k} \nonumber \\
&=& X^j \frac{\partial}{\partial q^j} + X^{k'} \frac{\partial^2 \Phi(q^i)}{ \partial q^k \partial q^{k'}} \frac{\partial}{\partial p_k}, 
\end{eqnarray}
where we have used the fact that $p_k = \frac{\partial \Phi(q^i)}{\partial q^k}$ and $i \in \{1,2,\cdots,n\}$ with $j \in J$ and $k, k' \in K$ such that $J \cup K = \{1,2,\cdots,n\}$ and $J \cap K = \{\}$. The map is a diffeomorphism if the Hessian $\frac{\partial^2 \Phi(q^i)}{ \partial q^k \partial q^{k'}}$ is nonsingular. Thus, the dynamical vector field $X$ on $T\mathcal{E}$ maps to a corresponding vector field $\psi_*X$ on $T \overline{\mathcal{E}}$. Notice that the vector field $\psi_*X$ which is tangent to the new Lagrangian submanifold $\overline{\mathcal{E}}$ may be generated from the same Hamiltonian (\ref{Hchoicegeneral}) by computing the Hamilton's equations for the variables $(q^j, p_k)$; this is because the Legendre transform can be interpreted as a canonical transformation. Thermodynamic-consistency relations are also fulfilled on $\overline{\mathcal{E}}$ and because of this, $H|_{\overline{\mathcal{E}}} = \Lambda$, a constant. 

\subsection{Relationship with the contact-geometric framework}
In the classical contact-geometric approach to equilibrium thermodynamics \cite{herm,mrug,old1,old2,old20,old22,old3,new4,new6,new7,new11}, one starts with a $(2n+1)$-dimensional contact manifold which is the pair $(\mathcal{C},\Theta)$, where $\mathcal{C}$ is a smooth manifold and $\Theta$ is a one-form satisfying $\Theta \wedge (d\Theta)^n \neq 0$. This means the hyperplane distribution ${\rm ker}(\Theta)$ is maximally non-integrable in the Frobenius sense and that the tangent bundle admits the Whitney-sum decomposition $T\mathcal{C} = {\rm ker}(\Theta) \oplus {\rm ker}(d\Theta)$, with ${\rm ker}(\Theta)$ being of codimension one while ${\rm ker}(d\Theta)$ is one-dimensional distribution generated by a vector field $\xi$, called the Reeb vector field, i.e., $\Theta(\xi) = 1$ and $d\Theta(\xi,\cdot) = 0$. From Darboux's theorem, the condition $\Theta \wedge (d\Theta)^n \neq 0$ implies that locally, there is a system of coordinates $(s,q^i,p_i)$, such that 
\begin{equation}
\Theta = ds - p_i dq^i, \quad \quad \xi = \frac{\partial}{\partial s}. 
\end{equation}
Hamiltonian dynamics is described by the following combined conditions:
\begin{equation}
\iota_{X_h} d\Theta = dh - \xi(h) \Theta,\quad \quad \Theta(X_h) = -h,
\end{equation} where\footnote{Let us denote with a lowercase $h$, a Hamiltonian function on a contact manifold.} $ h \in C^\infty(\mathcal{C}, \mathbb{R})$. These conditions imply that $X_h(h) = - h \xi(h)$, meaning that $h$ is conserved only in the level set $h^{-1}(0) \subset \mathcal{C}$, excluding the trivial case $\xi(h) = 0$. The counterparts of Lagrangian submanifolds in the contact-geometric setting are the maximal-dimensional submanifolds  $N \subset \mathcal{C}$ such that $\psi: N \hookrightarrow \mathcal{C}$ gives $\psi^* \Theta = 0$; equivalently, ${\rm dim}(N) = n$. These are called the Legendre submanifolds and from the local-coordinate expression $\Theta = ds - p_i dq^i$, one finds the local structure
\begin{equation}\label{Legendrelocal}
s = \mathcal{F}(q^i), \quad \quad d\mathcal{F}(q^i) - p_i dq^i = 0 \quad \implies \quad p_i = \frac{\partial \mathcal{F}(q^i)}{\partial q^i}, 
\end{equation}
similar to (\ref{Lagrangianlocal}) and (\ref{Firstlawgeneral}) for a Lagrangian submanifold of a symplectic manifold. This local resemblance between Lagrangian submanifolds of a symplectic manifold and Legendre submanifolds of a contact manifold immediately allows one to derive the thermodynamic counterparts of the results known from the contact-geometric framework of thermodynamics. In particular, the Theorems (\ref{theorem1}) and (\ref{theorem2}) are the Lagrangian counterparts of the corresponding results derived for Legendre submanifolds in the contact-geometric framework in \cite{new6} and \cite{new4}, respectively. In both the frameworks, the potentials are the generating functions. Our convention is that the Legendre transforms act as canonical transformations of the ambient symplectic phase space and therefore map $\mathcal{E}$, a Lagrangian submanifold, to a distinct embedded Lagrangian $\overline{\mathcal{E}}$; convexity then means precisely that these are diffeomorphic, i.e., ensemble equivalence is expressed as diffeomorphic equivalence of the corresponding Lagrangian embeddings. The symplectic and contact-geometric frameworks are, in fact, deeply connected. Taking the $2n$-dimensional symplectic phase space $(\mathcal{M},\omega)$ to be exact, i.e., $\omega = -d\theta$, the $(2n+1)$-dimensional contact phase space $(\mathcal{C},\Theta)$ can (at least, locally) be expressed as $\mathcal{C} = \mathcal{M} \times \mathbb{R}$. The inclusion map $\rho: \mathcal{M} \hookrightarrow \mathcal{C}$ gives $\rho^* \Theta = -\theta$.

\vspace{2mm}

Conceptually, the contact-geometric framework follows more naturally from statistical-mechanical arguments \cite{old1}, and one can treat the thermodynamically-conjugate variables together with the potential as distinct local coordinates of the contact phase space. Only when one is looking at equilibrium states, i.e., states satisfying $n+1$ constitutive relations in the form (\ref{Legendrelocal}) does one get Legendre submanifolds which are equivalent to what we have called spaces of equilibrium states in this paper. In contrast, the symplectic phase space incorporates only the thermodynamically-conjugate variables as its local coordinates and the space of equilibrium states is reached by the imposition of $n$ constitutive relations (\ref{Lagrangianlocal}). Another difference between the contact-geometric and symplectic frameworks is in the choice of the Hamiltonian function to describe a desired transformation. Since in the symplectic approach, the dynamics is conservative, it suffices to choose a Hamiltonian function that assumes any constant value (say, $\Lambda$) on the space of equilibrium states, i.e., $\mathcal{E} \subset H^{-1}(\Lambda)$. In the contact-geometric approach, however, due to the non-conservative nature of the Hamiltonian flow outside the level set $h^{-1}(0)$, the space of equilibrium states must be chosen to be contained within the level set $h^{-1}(0)$, i.e., the contact Hamiltonian must vanish for an equilibrium state and not assume any non-zero value. Let us end this discussion by noting the following well-known facts which can be derived in a straightforward manner \cite{Godbillon}:
\begin{enumerate}
\item Consider a $2n$-dimensional symplectic manifold $(\mathcal{M}, \omega)$ that is exact, i.e., $\omega = -d\theta$, where there is a Liouville vector field $\Delta$ satisfying $\iota_\Delta \omega = -\theta$. Then, if $\Lambda$ be a regular value of $H \in C^\infty(\mathcal{M}, \mathbb{R})$, the smooth level set $H^{-1}(\Lambda)$ is a $(2n-1)$-dimensional contact manifold if $\Theta = -\theta|_{H^{-1}(\Lambda)}$ is a contact one-form on $H^{-1}(\Lambda)$. This happens if $\Delta$ is transverse to $H^{-1}(\Lambda)$.
\item Consider a $(2n+1)$-dimensional contact manifold $(\mathcal{C}, \Theta)$ with Reeb vector field $\xi$. Let $h \in C^\infty(\mathcal{C},\mathbb{R})$ be a contact Hamiltonian function and $h=0$ be a regular value. Then the smooth level set $h^{-1}(0)$ is a $2n$-dimensional symplectic manifold with $\omega = -d\Theta|_{h^{-1}(0)}$, provided $\xi(h) \neq 0$ on $h^{-1}(0)$. The latter condition implies that $\xi$ is transverse to $h^{-1}(0)$, ensuring that $d\Theta|_{h^{-1}(0)}$ is non-degenerate.
\end{enumerate}

\section{Thermodynamic processes of the ideal gas}
Consider the ideal gas which satisfies the first law of thermodynamics as given by (\ref{gibbs1}) for the fundamental equation $E = E(S,V,N)$. Thus, the variables $q^i = (S,V,N)$ lie on the space of equilibrium states denoted by $\mathcal{E}$ whereas $p_i = (T,-P,\mu)$ are their thermodynamic conjugates (the so-called momenta). In all examples, we will take $k_B = 1$. 

\subsection{Isochoric process of the ideal gas}
Let us consider the following Hamiltonian:
\begin{equation}\label{Hamchoice1}
H = TS + \mu N - \gamma E(S,V,N) + \Lambda,
\end{equation} where $\gamma = (C+1)/C$ is the ratio of specific (per particle) heats of the ideal gas and $\Lambda$ is some real constant. Notice that the internal energy appears here as a function of the independent variables $(S,V,N)$. The Hamilton's equations (\ref{eom}) imply that the variables $(S,V,N)$ satisfy the following equations of motion:
\begin{equation}\label{set1}
\dot{S} = S, \quad \quad \dot{V} = 0, \quad \quad \dot{N} = N,
\end{equation} while the corresponding conjugate variables, i.e., $(T,-P,\mu)$, evolve as
\begin{equation}\label{eomnew111}
\dot{T} = -T + \gamma \frac{\partial E(S,V,N)}{\partial S}, \quad\quad \dot{P} = - \gamma \frac{\partial E(S,V,N)}{\partial V}, \quad\quad \dot{\mu} = -\mu + \gamma  \frac{\partial E(S,V,N)}{\partial N}. 
\end{equation}
On the space of equilibrium states on which the first law of thermodynamics (\ref{gibbs1}) holds, we find from (\ref{eomnew111}) and upon using $\gamma = 1 + (1/C)$ that
\begin{equation}\label{eomnew1112}
\dot{T} = \frac{T}{C}, \quad\quad \dot{P} = \gamma P, \quad\quad \dot{\mu} = \frac{\mu}{C}. 
\end{equation}
Integrating equations (\ref{set1}) and (\ref{eomnew1112}), one finds that (for $t \in \mathbb{R}^+$)
\begin{equation}\label{isochoricevolutionequations}
S(t) = S_0 e^t, \quad V(t) =  V_0, \quad N(t) = N_0e^t, \quad T(t) = T_0 e^{t/C}, \quad P(t) = P_0 e^{\gamma t}, \quad \mu(t) = \mu_0 e^{t/C}. 
\end{equation}
Here, $(S_0,V_0,N_0,T_0,P_0,\mu_0)$ are the values of the thermodynamic variables at $t = 0$. Clearly, the transformation is isochoric and the basic thermodynamic equations concerning the ideal gas are preserved. For example, $P(t)V(t) - N(t)T(t) = (P_0 V_0 - N_0T_0)e^{\gamma t}$, meaning that if the initial point corresponds to an equilibrium state of the ideal gas, i.e., it satisfies $P_0V_0 = N_0 T_0$, then the subsequent points on the phase trajectory are also equilibrium states of the ideal gas satisfying $P(t) V(t) = N(t) T(t)$. 

\vspace{2mm}

In order to determine how the thermodynamic potential (the internal energy $E(S,V,N)$) evolves as a function of $t$, we must resort to statistical mechanics. In fact, the well-known Sackur-Tetrode equation \cite{ST}, which describes the entropy of an ideal gas as a function of energy, volume, and number of particles in the thermodynamic limit may be used to express the internal energy of the gas as a function of entropy, volume, and number of particles. The result takes the form 
\begin{equation}\label{Eideal}
E(S,V,N) = A \exp[S/CN] V^{-1/C} N^{1 + 1/C},
\end{equation} where $A$ is a suitable positive constant. Substituting $S(t)$, $V(t)$, and $N(t)$ into the expression (\ref{Eideal}), one obtains
\begin{equation}
E(t) = E_0 e^{\gamma t},
\end{equation} where $E_0 = \big[A \exp[S_0/CN_0] V_0^{-1/C} N_0^{1 + 1/C}\big]$ and which is exactly consistent with the equipartition theorem since $E(t) = C N(t) T(t) = C N_0 T_0 e^{\gamma t} \sim e^{\gamma t}$. Thus, the evolution described by the Hamiltonian (\ref{Hamchoice1}) describes an isochoric transformation of the ideal gas in such a way that once we pick an initial point on the phase trajectory to be an equilibrium state of the ideal gas, all the subsequent points on the trajectory are equilibrium states preserving the thermodynamic equations consistently at each slice of $t$. 

\vspace{2mm}

The fact that an initial equilibrium point yields a trajectory confined to $\mathcal{E}$ follows from the constancy of $H$ on $\mathcal{E}$, together with the preservation of the constitutive relations which ensures that the Hamiltonian vector field is tangent to $\mathcal{E}$. Indeed, because the internal energy of the ideal gas is homogeneous of degree one in its arguments, Euler's theorem upon using (\ref{gibbs1}) gives the so-called Euler formula that reads $E(S,V,N) = TS - PV + \mu N$. Thus, for equilibrium states of the ideal gas, i.e., for points on $\mathcal{E}$ for which $\gamma E(S,V,N) = E(S,V,N) + PV$, (\ref{Hamchoice1}) gives $H|_\mathcal{E} = TS - PV + \mu N  - E(S,V,N) + \Lambda =\Lambda$, a constant, i.e., $\mathcal{E}$ is contained within the level set with constant $H$ and any phase flow passing through a point on $\mathcal{E}$ stays confined within $\mathcal{E}$, therefore preserving the equilibrium relations. 

\subsection{Isothermal-isochoric process of the ideal gas}
Let us now consider the following Hamiltonian:
\begin{equation}\label{Hchoice2}
H = TS - NT + \mu N - E(S,V,N) + \Lambda,
\end{equation} where $\Lambda$ is a real constant. Since for an ideal gas at equilibrium, one must have $PV = NT$, the Hamiltonian upon being restricted to the space of equilibrium states $\mathcal{E}$ turns out to be $H|_\mathcal{E} = TS - PV + \mu N - E(S,V,N) + \Lambda = \Lambda$, due to the Euler formula. Thus, the Hamiltonian assumes a constant value on the space of equilibrium states, ensuring its invariance under the dynamics induced by $H$. 

\vspace{2mm} 

The equations of motion (\ref{eom}) for the thermodynamic variables $(S,V,N)$ are given by
\begin{equation}\label{eomtherm0}
\dot{S} = S - N,  \quad \quad \dot{V} = 0, \quad \quad \dot{N} = N.
\end{equation}
Similarly, the dynamics of the variables $(T,-P,\mu)$ can be described from (\ref{eom}) as
\begin{equation}\label{eomtherm1}
\dot{T} = -T + \frac{\partial E(S,V,N)}{\partial S},  \quad \quad \dot{P} = - \frac{\partial E(S,V,N)}{\partial V}, \quad \quad \dot{\mu} = T - \mu + \frac{\partial E(S,V,N)}{\partial N}.
\end{equation}
Since we are interested in equilibrium states, we must restrict our attention to the space of equilibrium states on which the first law of thermodynamics (\ref{gibbs1}) holds, implying from (\ref{eomtherm1}) that
\begin{equation}\label{eomtherm2}
\dot{T} = 0,  \quad \quad \dot{P} = P, \quad \quad \dot{\mu} = T.
\end{equation}
Integrating (\ref{eomtherm0}) and (\ref{eomtherm2}), one finds that the thermodynamic variables evolve as (for $t \in \mathbb{R}^+$)
\begin{equation}\label{secondexampleevolution}
S(t) = (S_0 - N_0t)e^t, \quad V(t) = V_0, \quad N(t) = N_0e^t, \quad T(t) = T_0, \quad P(t) = P_0e^t, \quad \mu(t) = \mu_0 + T_0 t.
\end{equation}
Clearly, the thermodynamic transformation is both isothermal and isochoric. Given that an initial point  $(S_0,V_0,N_0,T_0,P_0,\mu_0)$ is suitably chosen so that the equilibrium relations of the ideal gas are satisfied (e.g., $P_0V_0 = N_0 T_0$), subsequent points are all equilibrium states of the ideal gas. For instance, we have $P(t) V(t) - N(t) T(t) = (P_0 V_0 - N_0 T_0)e^t = 0$. The corresponding evolution of the internal energy is obtained by substituting $S(t)$, $V(t)$, and $N(t)$ into (\ref{Eideal}) which gives
 \begin{equation}\label{Eideal2222}
E(t) = E_0 e^t,
\end{equation} where $E_0 = \big[A \exp[S_0/CN_0] V_0^{-1/C} N_0^{1 + 1/C}\big]$. This is consistent with the equipartition theorem since $E(t) = C N(t) T(t) = C N_0 T_0 e^t \sim e^t$. 

\section{Infinitesimal transformations between systems}\label{mapsec}
So far we have dealt with situations where a suitably-chosen Hamiltonian can describe a thermodynamic process of a system. The essential idea has been to choose a Hamiltonian such that on the space of equilibrium states, it assumes a constant value so that the space of equilibrium states is invariant under the corresponding Hamiltonian flow. Let us now consider situations where the Hamiltonian is not constant on the space of equilibrium states. Quite naturally, the corresponding dynamics will not be a thermodynamic process of the concerned system. For the ideal gas in the microcanonical ensemble with the first law (\ref{gibbs1}), let us take the following Hamiltonian:
\begin{equation}
H = \frac{\alpha_0 N^2}{V}, 
\end{equation} where $\alpha_0 > 0$ is a constant. The corresponding Hamilton's equations turn out to be 
\begin{eqnarray}\label{eqnm1twobodyint}
\dot{S} = 0, \quad \quad \dot{V} = 0, \quad \quad \dot{N} = 0, \quad \quad \dot{T} = 0, \quad \quad \dot{P} = - \frac{\alpha_0 N^2}{V^2}, \quad \quad \dot{\mu} = - \frac{2\alpha_0 N}{V}. 
\end{eqnarray}
The resulting evolution does not preserve the thermodynamic (equilibrium) relations of the ideal gas even if the initial point corresponds to an equilibrium state of the ideal gas. In other words, even if $P_0 V_0 = N_0 T_0$, one finds $P(t) V(t) \neq N(t) T(t)$ for $ t > 0$. However, it is interesting to note that the pressure of the system evolves as (for $t \in \mathbb{R}^+$)
\begin{equation}\label{Pttwobodyint}
P(t) = P_0 - \frac{\alpha_0 N_0^2}{V_0^2} t.
\end{equation}
Thus, choosing $P_0$ to be the pressure of the ideal gas, i.e., $P_0 = N_0 T_0/V_0 = N(t)T(t)/V(t)$, the evolving pressure turns out to be 
\begin{equation}\label{twobodyeqnofstate}
P(t) = \frac{N_0 T_0}{V_0} - \frac{\alpha_0 N_0^2}{V_0^2} t , 
\end{equation} i.e., the Hamiltonian flow maps the ideal gas to a one-parameter family of interacting gases with two-body interactions. More precisely, at each constant slice of $t$, one gets an interacting gas with two-body interactions being characterized by the constant $\alpha_0 t$, resembling the van der Waals model which has been discussed in \cite{old22}. Put differently, one starts with a point on the space of equilibrium states of the ideal gas, say, $\mathcal{E}_{\rm ideal}$ and then as the flow progresses, the trajectory passes transversely through a one-parameter family of spaces of equilibrium states, say, $\mathcal{E}_{\rm interacting}^t$. The above equation is closely related to the van der Waals model; taking the van der Waals equation ($a,b > 0$) $P = \frac{NT}{V - Nb} - \frac{aN^2}{V^2}$, we get for $V \gg Nb$, the following approximate result:
\begin{equation}
P = \frac{NT}{V} \bigg(1 - \frac{Nb}{V}\bigg)^{-1} - \frac{aN^2}{V^2} \approx \frac{NT}{V} - \frac{N^2(a - Tb)}{V^2}.
\end{equation} This has a similar structure as (\ref{twobodyeqnofstate}). Thus, one can make use of Hamiltonian flows to describe maps between related thermodynamic systems. In that case, one must pick the Hamiltonian function to be such that it does not assume a constant value on the space of equilibrium states of the system one starts with. One can construct multiple examples of this type upon using Hamiltonian dynamics on symplectic manifolds without resorting to contact Hamiltonian dynamics like in \cite{old2,old22,new7}. Let us consider two novel examples which have not appeared in the literature on geometrical thermodynamics so far. 

\subsection{Interacting gas with two-body and four-body interactions}\label{24body}
In order to map the ideal gas to a gas with two-body and four-body interactions, the following could be a plausible choice of the Hamiltonian:
\begin{equation}
H =  \frac{\alpha_0 N^2}{V} + \frac{\beta_0 N^4}{3V^3}, 
\end{equation} where $\alpha_0,\beta_0 \in \mathbb{R}$ are constants whose sign should be chosen so as to correspond to attractive or repulsive interactions (in each term), as the case may be. For this case, taking the initial point to be an equilibrium state of the ideal gas, i.e., for $P_0V_0 = N_0T_0$, the pressure evolves as
\begin{eqnarray}
P(t) &=& \frac{N_0 T_0}{V_0} - \bigg(\frac{\alpha_0 N_0^2}{V_0^2}  + \frac{\beta_0 N_0^4}{V_0^4} \bigg) t, \quad \quad t \in \mathbb{R}^+,
\end{eqnarray}
where in typical situations, one would have\footnote{Corresponding to effective two-body attraction and four-body repulsion.} $\alpha_0 > 0$ and $\beta_0 < 0$. This gives a one-parameter family of interacting gases related to the ideal (non-interacting) gas via a Hamiltonian flow. 

\subsection{Redlich-Kwong equation}\label{RKsec}
The Redlich-Kwong model is a two-parameter real-gas model which is often more accurate than the van der Waals equation and some real-gas models with more than two parameters. The equation of state is given by
\begin{equation}\label{RKM}
  \bigg(P + \frac{\alpha_0 N^2}{\sqrt{T}V(V+N\beta_0)}\bigg)(V-N\beta_0) = NT,
\end{equation} where $\alpha_0$ and $\beta_0$ are suitable constants. Consider a pair of Hamiltonians on the phase space with coordinates $q^i = (S,V,N)$ and $p_i = (T, -P, \mu)$ as given by 
\begin{equation}
H_1 = -\beta_0 NP, \quad \quad H_2 = -\frac{\alpha_0 N}{\sqrt{T}\beta_0} \ln \bigg(\frac{V}{V+N\beta_0}\bigg). 
\end{equation}
It is easy to verify that $X_{H_1} (P) = 0$ and $X_{H_2} (V) = 0$, while $X_{H_{1,2}} (N) = X_{H_{1,2}} (T) = 0$. Moreover, one has
\begin{equation}
X_{H_1} (V) = \beta_0 N, \quad \quad X_{H_2} (P) = - \frac{\alpha_0 N^2}{\sqrt{T} V (V + N \beta_0)}.
\end{equation}
That is, the volume deforms under the flow of $X_{H_1}$ while the pressure deforms under the flow of $X_{H_2}$. Taking the initial conditions $(P_0, V_0, N_0, T_0)$, one therefore gets
\begin{equation}
V(t_1) = V_0 + \beta_0 N_0 t_1, \quad \quad P(t_2) = P_0 - \frac{\alpha_0 N_0^2}{\sqrt{T_0}V_0(V_0+\beta_0 N_0)} t_2,
\end{equation} where $t_1, t_2 \in \mathbb{R}^+$ are the affine parameters that parametrize the integral curves of $X_{H_1}$ and $X_{H_2}$, respectively. It thus turns out that a successive application of $X_{H_1}$ followed by $X_{H_2}$ gives
\begin{equation}\label{deformedequationRKM111}
P_0 V_0 =  \bigg(P(t_2) + \frac{\alpha_0 N_0^2}{\sqrt{T_0}V(t_1)(V(t_1)+N_0\beta_0)} t_2\bigg)(V(t_1)-N_0\beta_0 t_1) 
\end{equation}
Here, the $V_0$ is replaced by $V(t_1)$ in the second term within the first parenthesis of the right-hand side is due to the dynamics of $X_{H_1}$ which has been applied before $X_{H_2}$, i.e., $V(t_1)$ is the initial volume under the flow of $X_{H_2}$ which preserves the volume as discussed above. Considering the initial point to be an equilibrium state of the ideal gas, i.e., $P_0V_0 = N_0 T_0$, (\ref{deformedequationRKM111}) gives 
\begin{equation}\label{RKM1}
  \bigg(P(t_2) + \frac{\alpha_0 N_0^2}{\sqrt{T_0}V(t_1)(V(t_1)+N_0\beta_0)} t_2\bigg)(V(t_1)-N_0\beta_0 t_1) = N_0T_0,
\end{equation} which coincides with a family of Redlich-Kwong-type equations. Notice that the vector fields $X_{H_1}$ and $X_{H_2}$ do not commute, i.e., their Lie bracket $[X_{H_1}, X_{H_2}] \neq 0$. Consequently, starting with the ideal gas with equation $P_0 V_0 = N_0 T_0$, a successive application of $X_{H_2}$ followed by $X_{H_1}$ gives a different equation of state which reads
\begin{equation}\label{RKM2}
  \bigg(P(t_2) + \frac{\alpha_0 N_0^2}{\sqrt{T_0}(V(t_1) - \beta_0 N_0 t_1)[V(t_1)+N_0\beta_0 (1 - t_1) ]}t_2\bigg)(V(t_1)-N_0\beta_0 t_1) = N_0T_0.
\end{equation} Note that both equations (\ref{RKM1}) and (\ref{RKM2}) describe two-parameter families of thermodynamic systems. In (\ref{RKM2}), the volume gets deformed in the second term within the first parenthesis of the left-hand side due to the dynamics of $X_{H_1}$ which has been applied after the deformation of the pressure has been achieved due to the flow described by $X_{H_2}$.

\section{An irreversible transformation: Free expansion}
So far we have observed that Hamiltonian dynamics can be effectively employed to describe either thermodynamic processes of a system or infinitesimal transformations connecting different systems. We shall now demonstrate via a concrete example that the Hamiltonian approach can describe irreversible transformations as well. As a concrete example, let us consider the free expansion of the ideal gas into vacuum, from an initial volume $V_i$ to a final volume $V_f$. As is well known \cite{ST}, the gas performs no work in the expansion, acquires/loses no heat, and by the first law of thermodynamics, the internal energy also does not change. Since we are assuming that the constitutive relations of the ideal gas hold during the process, $E = CNT$ implies that the temperature remains constant. Nevertheless, this transformation leads to a finite and irreversible change of the entropy of the ideal gas, as given by \cite{ST}
\begin{equation}\label{DeltaSfree}
\Delta S = N \ln \bigg( \frac{V_f}{V_i}\bigg). 
\end{equation}
In order to describe the dynamics in a Hamiltonian framework, let us consider the first law of thermodynamics with fixed $N$ so that it simplifies to $dE = TdS - PdV$. That is, the symplectic phase space is four-dimensional with $(q^i,p_i) = \{(S,T), (V,-P)\}$ and the fundamental equation $E=E(S,V)$ is a function on the two-dimensional Lagrangian submanifold representing the space of equilibrium states. Let us now take a Hamiltonian of the form
\begin{equation}
H = - \kappa \frac{PV}{T}, 
\end{equation} where $\kappa > 0$ is a real constant. Then the Hamiltonian vector field is of the following form:
\begin{equation}
X_H = \frac{ \kappa PV}{T^2} \frac{\partial}{\partial S} + \frac{ \kappa V}{T} \frac{\partial}{\partial V} -  \frac{ \kappa P}{T} \frac{\partial}{\partial P}. 
\end{equation}
As a result, we have the equations of motion
\begin{equation}
\dot{S} =  \frac{ \kappa PV}{T^2}, \quad \quad \dot{T} = 0,  \quad \quad \dot{V} =  \frac{ \kappa V}{T}, \quad \quad \dot{P} = - \frac{ \kappa P}{T},
\end{equation} which are solved to give
\begin{equation}\label{evolutionfreeexp}
S(t) =S_0 + \bigg( \frac{ \kappa P_0V_0}{T_0^2}\bigg) t, \quad \quad T(t) = T_0, \quad \quad V(t) = V_0\exp[ (\kappa/T_0) t] , \quad \quad P(t) = P_0 \exp[-(\kappa/T_0) t]. 
\end{equation}
Clearly, $P(t) V(t) = P_0 V_0 = N T_0$, i.e., the ideal-gas equation is preserved for all $t$ provided that the initial point (subscripts `0') is an equilibrium state of the ideal gas. Moreover, on the space of equilibrium states, we can write $H = -\kappa N$ due to the ideal gas equation which is a constant. In other words, the Lagrangian submanifold corresponding to the space of equilibrium states of the ideal gas is invariant under the flow of $X_H$, guaranteeing that all points along the trajectory are equilibrium states of the ideal gas, provided $P_0V_0 = N T_0$; this follows from $X_H(PV - NT) = 0$ if $P_0V_0 = N T_0$. Consider the initial volume to be $V_0 = V_i$ and the volume after an instant of time $t=\tau$ to be $V(\tau) = V_f$. Then $V_f = V_i \exp[ (\kappa/T_0) \tau]$ or $(\kappa/T_0) \tau = \ln (V_f/V_i)$. Substituting this into the expression for $S(t)|_{t=\tau}$ from (\ref{evolutionfreeexp}) gives us
\begin{equation}
S(\tau) = S_0 + N \ln \bigg( \frac{V_f}{V_i} \bigg),
\end{equation} which agrees with the result (\ref{DeltaSfree}). It may be noted that while the physical free expansion is an irreversible process, the Hamiltonian flow
constructed above furnishes a reversible path lying entirely within the space of equilibrium states that connects the same initial and final states and reproduces the correct
entropy change.

\section{Port-Hamiltonian framework for thermodynamics}
As a final application of the theory of Hamiltonian systems to thermodynamics, we shall now present a port-Hamiltonian framework. In generic port-Hamiltonian framework \cite{port1,port2,port3,port4} (see also, \cite{new50,new80}), the basic idea is to supplement the conservative Hamiltonian structure with (a) input/output ports representing input/output of energy, and (b) employ the Poisson structure generalizing the symplectic structure. Let us restrict ourselves to the symplectic structure and propose the addition of input/output ports. 

\begin{axiom}
Let $(\mathcal{M},\omega)$ be a symplectic manifold and $H\in C^\infty(\mathcal{M},\mathbb{R})$ the Hamiltonian function. Let $\{Y_i\}_{i=1}^m$ be vector fields on $\mathcal{M}$ (also known as the port vector fields), together with the so-called scalar inputs $u_i(t)$ and outputs $y_i(t) = \iota_{Y_i} dH$. A symplectic port-Hamiltonian system is then defined by
\begin{equation}
\label{sph}
X = X_H + \sum_{i=1}^m u_i Y_i .
\end{equation}
\end{axiom}
\textbf{Remark 1:} The key structural property of this construction is the `power balance' as given by the condition
\begin{equation}
\label{powerbalance}
X(H) = \sum_{i=1}^m u_i y_i .
\end{equation} 

\noindent
\textbf{Remark 2:} Setting $u_i = 0$ gives back the original form of Hamiltonian dynamics. In other words, the ports exchange power with the system, and the power-balance condition (\ref{powerbalance}) is the analogue of the first law of thermodynamics. Irreversibility and dissipation can now arise in this setting, depending upon how one chooses $\{u_i\}$. In the following examples, we shall restrict the port-Hamiltonian dynamics to the equilibrium Lagrangian submanifold within the symplectic phase space so that we can take the internal energy as the Hamiltonian and determine the intensive variables from the constitutive relations.

\subsection{Isothermal expansion of the ideal gas}
We shall now demonstrate via a simple example, the application of the port-Hamiltonian framework to the thermodynamics of the ideal gas. Let us consider the ideal gas in the fixed-particle scenario discussed in the case of free expansion; $N$ is a mere parameter and the symplectic phase space is four-dimensional, with the symplectic two-form $\omega=dS \wedge dT - dV \wedge dP$. A natural choice of the Hamiltonian would be $H = E(S,V)$. In order to describe isothermal expansion of the ideal gas against a piston, let us introduce the port vector fields
\begin{equation}\label{SV_portvectorfields}
Y_S = \frac{\partial}{\partial S}, \quad \quad Y_V = \frac{\partial}{\partial V},
\end{equation} inputs $u_S,u_V$, and outputs (at equilibrium) $y_S=T$, $y_V=-P$. Then the power-balance condition (\ref{powerbalance}) amounts to
\begin{equation}\label{portidealeqnbalance}
\dot{E} = u_S T - u_V P. 
\end{equation}
Choosing $u_S = \dot{S}$ and $u_V = \dot{V}$ gives us the first law of thermodynamics $dE = TdS - PdV$. Since in the case of isothermal expansion, the ideal gas expands against a piston by drawing heat from a thermal reservoir, let us introduce two interconnections: 
\begin{enumerate}
\item \textbf{Mechanical port:} A piston subject to an external pressure $P_{\rm ext}(t)$ and with a linear damping $\gamma_m$, such that
\begin{equation}
P - P_{\rm ext}(t) = \gamma_m \dot V.
\end{equation} 

\item \textbf{Thermal port:} A heat bath at temperature $T_{\rm ext}$. Assuming perfect and instantaneous transfer of energy to the gas, we can take $y_S = T = T_{\rm ext}$.

\end{enumerate}

\noindent
On the space of equilibrium states, one has
\begin{equation}
\label{isomanifold}
 P=\frac{NT_{\rm ext}}{V}, \quad \quad dS = N  \frac{dV}{V}.
\end{equation}
Thus, combining $P=\frac{NT_{\rm ext}}{V}$ with $P-P_{\rm ext}=\gamma_m\dot V$, we get the result
\begin{equation}\label{dotVlambert}
 \dot{V} = \frac{1}{\gamma_m}\!\left(\frac{NT_{\rm ext}}{V}-P_{\rm ext}(t)\right).
\end{equation}
Although the above differential equation cannot, in general, be solved analytically, taking the pressure to be time-independent, i.e., $P_{\rm ext}(t) = P_0$, it can be solved using the Lambert $W$ function \cite{W}. For convenience, let us define the positive constants
\begin{equation}
\mathcal{A} = \frac{N T_{\rm ext}}{\gamma_m},\quad \quad \mathcal{B} = \frac{P_0}{\gamma_m},
\end{equation}
and impose the initial condition $V(0)=V_0$. Writing $U_0 = \mathcal{A} - \mathcal{B} V_0$, the solution $V(t)$ of (\ref{dotVlambert}) is of the following form:
\begin{equation}
V(t) = \frac{\mathcal{A}}{\mathcal{B}}\left[ 1 + W\!\left(- \frac{U_0}{\mathcal{A}} 
\exp\!\left\{-\frac{(U_0 + \mathcal{B}^2 t)}{\mathcal{A}}\right\}\right)\right],
\end{equation}
where $W(\cdot)$ is the Lambert $W$ function; for the monotone relaxation to $V=\frac{\mathcal{A}}{\mathcal{B}}=\frac{N T_{\rm ext}}{P_0}$, one must take the principal branch of $W$. As a final step, let us compute the heat input from the heat bath during the course of the isothermal expansion. Because the temperature stays constant, $\dot{E} = 0$, which is the same as saying that $X(H) = 0$. Using this in (\ref{portidealeqnbalance}), we can write $u_S T = u_V P$, or substituting for $u_S$ and $u_V$, the result
\begin{equation}
T_{\rm ext} \dot{S} =  P \dot{V}. 
\end{equation} 
Thus, the heat input is given by
\begin{equation}
Q = \int T_{\rm ext} \dot{S} dt = \int P \dot{V} dt. 
\end{equation}
Putting $P = P_{\rm ext} + \gamma_m \dot{V}$ as taken before, one finds the intriguing form
\begin{equation}
Q = \int P_{\rm ext} dV(t) + \gamma_m \int \dot{V}(t)^2 dt,
\end{equation} where we have made the time-dependencies explicit. The first term above is just the usual `$PdV$'-work while the second term accounts for possible dissipative losses due to friction on the piston. In the quasi-static (reversible) limit, the system passes through a continuous family of equilibrium states. At each instant, imposing $\dot V \approx 0$ in (\ref{dotVlambert}) yields the instantaneous equilibrium condition $P_{\rm ext}(t) V(t) = N T_{\rm ext}$. 

\subsection{Irreversible, isochoric heat transfer to an ideal gas}
Let us consider an ideal gas confined in a rigid vessel with constant volume $V = V_0$. Taking $N$ to be fixed and a mere parameter, the symplectic phase space is four-dimensional with the symplectic two-form $\omega = dS \wedge dT - dV \wedge dP$. We shall consider the situation where the gas exchanges heat irreversibly with a thermal reservoir at fixed 
temperature $T_{\rm ext}$ through a thermal conductor with conductance $K$. Let us work out the port-Hamiltonian framework for this transformation. In this case, let us take $u_S = \dot{S}$ and $u_V = 0$, the last choice being motivated by the constancy of volume (unlike the previous example where the volume could expand by pushing the piston). This gives us the power-balance condition to be
\begin{equation}
\dot{H} = T\dot{S} = T \Sigma, 
\end{equation} where $\Sigma$ is the system's entropy-production rate. Since the gas is connected to a heat bath at fixed temperature $T_{\rm ext}$, the connection between the gas and the bath is given by Fourier's law for heat flow through a thermal conductor with conductance $K$ in the manner
\begin{equation}
\dot{Q} = K (T_{\rm ext} - T) \quad \implies \quad \Sigma = \frac{\dot{Q}}{T} = \frac{K}{T}(T_{\rm ext} - T).
\end{equation}
The non-conservative nature of the dynamics is immediately apparent if we notice that $\dot{H} = T \Sigma = K( T_{\rm ext} - T) > 0$, since $T_{\rm ext} > T$ for net heat flow into the ideal gas. Taking $H$ to be the internal energy of the ideal gas, i.e., $H = E = (3/2) N T$, one can therefore get the condition
\begin{equation}
\frac{3N}{2} \dot{T} = K(T_{\rm ext} - T). 
\end{equation}
Solving for $T$, one obtains 
\begin{equation}
T(t) = T_{\rm ext} + (T_0 - T_{\rm ext}) \exp [-(2K/3N) t] ,
\end{equation} where $T(0) = T_0$. The ideal-gas equation can be satisfied for each $t$ with $P(t) = N T(t)/V_0$, since $V(t) = V_0$. In addition to quantifying the entropy production in the ideal gas, one can also quantify the total entropy production. The entropy-production rates of the system and the bath are given by
\begin{equation}
\dot{S}_{\rm gas} = \Sigma,
\quad \quad 
\dot{S}_{\rm bath} = -\frac{\dot Q}{T_{\rm ext}},
\end{equation} respectively. Thus, the total entropy-production rate of the universe is given by
\begin{equation}
 \dot{S}_{\rm gas} + \dot{S}_{\rm bath} 
= K (T_{\rm ext} - T)\!\left(\frac{1}{T} - \frac{1}{T_{\rm ext}}\right) \geq 0.
\end{equation}
The above-derived quantity is positive even if $T_{\rm ext} < T$, i.e., in situations where the ideal gas loses heat to the bath. This is consistent with the second law of thermodynamics because entropy of the universe must increase in any irreversible process, including heat transfer (finite $K$). In this example, we have assumed that the thermal conductor coupling the heat bath with the ideal gas does not have any dissipative losses. 

\section{Conclusions}\label{dissec}
In this paper, we have presented a self-contained symplectic framework suited for describing thermodynamics of equilibrium states in which the space of equilibrium states arises as a Lagrangian submanifold of the symplectic phase space and generated by the thermodynamic potential. The thermodynamic transformations are realized as Hamiltonian flows that preserve this submanifold. This viewpoint cleanly complements the familiar contact-geometric framework: it preserves the conservative geometric backbone, renders changes of thermodynamic ensemble as (partial) Legendre transforms between Lagrangian submanifolds, and yields explicit evolution laws through the standard Hamiltonian toolkit. The worked-out examples involving the ideal gas clearly demonstrate that the formalism reproduces the textbook relations \cite{ST} (e.g., Euler homogeneity, equipartition theorem, and entropy changes) while keeping the geometry transparent. The framework also accommodates mappings between different thermodynamic systems by choosing Hamiltonians that describe flows across families of Lagrangian submanifolds (e.g., towards van der Waals or Redlich-Kwong-type behavior starting from the ideal gas). The symplectic framework thus provides a clear and unified geometric language for thermodynamics, closely aligned with the familiar tools of classical mechanics.

\vspace{2mm}

It should be mentioned that one can also discuss black hole thermodynamics using our symplectic picture of thermodynamics wherein one would be able to construct maps (in the thermodynamic space) between black holes in different gravity theories as presented earlier in \cite{new7} using the contact-geometric setting. Beyond closed systems, we have shown that open thermodynamic processes fit naturally into a symplectic port-Hamiltonian template, with thermal and mechanical ports delivering the power balance by construction. The isothermal expansion against a piston illustrates how the port-Hamiltonian construction can be realized and how dissipation is cleanly separated from the useful work in the energy accounting, while the problem of heat transfer between a heat bath and an ideal gas via a thermal conductor illustrates the natural emergence of irreversibility in the port-Hamiltonian framework. Future work may explore how the port-Hamiltonian approach can be extended to model realistic irreversible processes and practical energy-exchange systems in thermodynamics.

\section*{Acknowledgements} We are thankful to A. P. Balachandran, C. Bhamidipati, and S. Chaturvedi for encouragement during the early stages of the work. A.G. gratefully acknowledges detailed discussions with P. Guha concerning port-Hamiltonian systems and also thanks him for pointing out some important references. A.G. acknowledges support from the Ministry of Education (MoE), Government of India in the form of a Prime Minister's Research Fellowship (ID: 1200454) and the School of Physics, University of Hyderabad for hospitality and local-travel support through the IoE-UoH-IPDF (EH) scheme. Our dear colleague A. P. Balachandran passed away on 18th April 2025, before this paper reached its final form. In the honor of his memory, we are dedicating this work.

\appendix

\section{Energy and entropy representations}\label{appA}
Consider a hydrostatic system with the first law of thermodynamics given by (\ref{gibbs1}) where $E = E(S,V,N)$ is the fundamental equation. Since $T > 0$, one can rewrite the first law of thermodynamics as
\begin{equation}\label{gibbsentropy}
dS = \frac{dE}{T} + \frac{P}{T} dV - \frac{\mu}{T} dN,
\end{equation} where $S = S(E,V,N)$ is the thermodynamic potential; this is obtained by solving for $E = E(S,V,N)$ in favor of $S$. Thus, we shall refer to (\ref{gibbs1}) as the energy representation while (\ref{gibbsentropy}) shall be referred to as the entropy representation (see also, \cite{new80}). Both representations contain the same physical information and are defined in the same ensemble, namely, the microcanonical ensemble. 

\section{Ensembles from hydrostatic systems}\label{appB}
If we revisit the case of a hydrostatic system where (\ref{gibbs1}) describes the first law of thermodynamics, upon defining $F(T,V,N) = E(S,V,N) - TS$ with $\frac{\partial^2 E}{\partial S^2} \neq 0$, one can write $dF = -S dT - PdV + \mu dN$ which is the first law of thermodynamics in the canonical ensemble. Similarly, defining $\mathcal{H}(S,P,N) = E(S,V,N) + PV$ with $\frac{\partial^2 E}{\partial V^2} \neq 0$ gives $d\mathcal{H} = TdS + VdP + \mu dN$, which is the first law of thermodynamics in the isoenthalpic-isobaric ensemble. Yet another commonly-encountered ensemble is the isothermal-isobaric ensemble which is achieved as $G(T,P,N) = E(S,V,N) + PV - TS = \mathcal{H}(S,P,N) - TS = F(T,V,N) + PV$, giving $dG = -SdT + VdP + \mu dN$. Thus, different statistical ensembles are connected by Legendre transforms in the thermodynamic limit, while within each ensemble one can describe multiple representations of the first law of thermodynamics as we already observed in the context of the energy and entropy representations of the microcanonical ensemble as given by (\ref{gibbs1}) and (\ref{gibbsentropy}), respectively.

\end{document}